\documentstyle[prl,aps]{revtex}
\input BoxedEPS.tex
\SetOzTeXEPSFSpecial
\HideDisplacementBoxes

\begin{document}

\twocolumn[\hsize\textwidth\columnwidth\hsize
           \csname @twocolumnfalse\endcsname

\title{Unconventional isotope effects in the high-temperature cuprate 
superconductors}
\author{Guo-meng~Zhao$^{1}$, H. Keller$^{1}$ and K. Conder$^{2}$} 
\address{$^{1}$Physik-Institut der Universit\"at Z\"urich,
CH-8057 Z\"urich, Switzerland\\
$^{2}$Laboratory for Neutron Scattering, ETH Z\"urich and PSI 
Villigen, CH-5232 Villigen PSI, Switzerland}

\maketitle
\widetext

\begin{abstract}
We review various isotope effects in the 
high-$T_c$ cuprate superconductors to assess the role of
the electron-phonon interaction in the basic physics of 
these materials. Of particular interest are the unconventional 
isotope effects on the supercarrier mass, on the charge-stripe 
formation temperature, on the pseudogap formation temperature, on 
the EPR linewidth, on the spin-glass freezing temperature, and on the 
antiferromagnetic ordering temperature. The observed unconventional 
isotope effects strongly suggest that lattice vibrations play an 
important role in the microscopic pairing mechanism of high-temperature 
superconductivity.

\end{abstract}
\vspace{1cm}
]
\narrowtext

\section{introduction}
Developing a correct microscopic theory for high-$T_c$ superconductivity is 
one of the most challenging problems in condensed matter physics.  
More than ten years after the discovery of the high-$T_c$ cuprate 
superconductors by 
Bednorz and M\"uller \cite{KAM86}, there have been no microscopic theories
that can describe the physics
of high-$T_c$ superconductors completely and unambiguously. Due to the 
high $T_c$ values and the earlier observation of a small oxygen-isotope effect 
in a 90 K cuprate superconductor YBa$_{2}$Cu$_{3}$O$_{7-y}$ 
\cite{Batlogg,Bourne87,Don88}, many theorists 
believe that the electron-phonon interaction cannot be the origin of 
high-$T_c$ superconductivity. Most physicists have thus turned their 
minds towards an alternative pairing interaction of purely electronic
origin (e.g., see Ref.\cite{Schrieffer,Millis,Anderson}). As a matter of fact, the idea that the highest $T_{c}$ is only 
30 K within the conventional phonon-mediated mechanism is not justified. 
In principle, $T_{c}$ can increase monotonically with both 
the phonon frequency and the electron-phonon coupling constant within 
the phonon-mediated Eliashberg theory \cite{CarbotteRev}. The limitation of the 
$T_{c}$ within the phonon mechanism is only imposed by 
a possible structural instability in the case 
of too strong an electron-phonon interaction. However, there is no 
universally accepted, simple, and quantitative stability criterion 
\cite{CarbotteRev}.

It is well known that the observation of a gap in the 
electronic excitation spectrum 
\cite{Daunt46} and the discovery of an isotope 
effect \cite{Maxwell50,Reynolds50} in conventional superconductors 
provided important and crucial clues to the understanding of the 
microscopic mechanism of superconductivity.
In particular, the effect of changing isotope mass on the 
superconducting transition 
temperature $T_c$ implies that superconductivity is not 
of purely electronic origin, but that lattice vibrations (phonons) 
play an important role
in the microscopic mechanism for this phenomenon.

The first evidence for an
isotope effect was reported in 1950 by Maxwell \cite{Maxwell50} and 
independently by Reynolds {\em et al.} \cite{Reynolds50}. 
They found that the critical temperature $T_c$ 
of mercury is an inverse function of the isotope mass.
In the same year Fr\"ohlich \cite{Froehlich50} pointed out that the 
same electron-lattice interaction which describes the scattering of 
conduction electrons by lattice vibrations gives rise to an indirect 
interaction between electrons. He proposed that this indirect interaction 
is responsible for superconductivity. Fr\"ohlich's theory got strong support 
from the observed isotope effect, and played a decisive role in 
establishing a correct mechanism. In 1956, Cooper \cite{Cooper56} 
demonstrated that electrons with an attractive 
interaction form bound pairs 
(so called Cooper pairs) which lead to superconductivity. 
However, the existence of electron pairs does not necessarily 
imply a phonon mediated 
pairing. Indeed, Bose condensation 
as considered in 1955 by Schafroth \cite{Schafroth55}, 
is also a possible mechanism for superconductivity, but the 
model was not able to explain the isotope effect. 
Finally, in 1957, Bardeen, Cooper and Schrieffer \cite{Bardeen57} 
developed the BCS theory which can explain most physical properties 
observed in conventional superconductors.

Remarkably, the BCS theory can well explain the isotope 
effect. The $T_{c}$ within the theory is given by
 \begin{equation}\label{Ie1}
 	k_B T_c = 1.13 \hbar \omega_{D} \exp{\left( -\frac{1}{N(0)V}\right) } ~,
 \end{equation}
where $\omega_{D}$ is the Debye 
frequency, which is proportional to $M^{-1/2}$. The electron-phonon coupling $N(0)V$ is the product 
of an electron-phonon interaction 
strength $V$ and the electronic density of states at the Fermi surface 
$N(0)$, both of which are independent of the ion mass $M$ in the 
harmonic approximation. Eq.~1 implies an isotope-mass dependence of $T_{c}$, with an 
isotope-effect exponent $\alpha = - d\ln T_{c}/d\ln M$ = 1/2. 
This is in excellent agreement with the
reported isotope effects  
in the non-transition metal superconductors (e.g., Hg, Sn and Pb). 
In fact, the isotope effect was 
the first justification of the proposed electron-phonon coupling 
mechanism.

The conventional phonon-mediated 
superconducting theory is based on the Migdal adiabatic approximation 
in which the phonon-induced electron self-energy is given correctly 
to the order of $(m_{b}/M)^{1/2}$ $\sim$ 10$^{-2}$, where $m_{b}$ is the 
bare mass of an 
electron. Within this approximation, 
the density of states at the Fermi level  $N(0)$, the electron-phonon coupling 
constant $\lambda_{ep}$, and the effective mass of the supercarriers are all 
independent of the ion mass $M$. However, if the interactions between 
electrons and nuclear ions are strong enough for electrons to form 
polarons (quasiparticles dressed by lattice distortions), their
effective mass  $m^{*}$ will depend on
$M$. This is because the polaron mass $m^{*}= m_{b} \exp (A/\omega)$
\cite{ale}, where $A$ is a constant, and $\omega$ is a characteristic
optical phonon frequency which depends on the masses of ions.  Hence, there
is a large isotope effect on the carrier mass in polaronic metals, in
contrast to the zero isotope effect in ordinary metals. The total exponent
of the isotope effect on $m^{*}$ is defined as $\beta  = \sum - d\ln
m^{*}/d\ln M_{i}$ ($M_{i}$ is the mass of the $i$th atom in a unit cell). From this definition and 
the expression for the polaron mass $m^{*}$ mentioned
above, one readily finds
\begin{equation}\label{Ie2}
\beta = -{1\over{2}}\ln(m^*/m_{b}).
\end{equation}
The above equation implies that there should be a large negative 
isotope effect on $m^{*}$ in polaronic metals.
\begin{figure}[htb]
\input{epsf}
\epsfxsize7cm
\centerline{\epsfbox{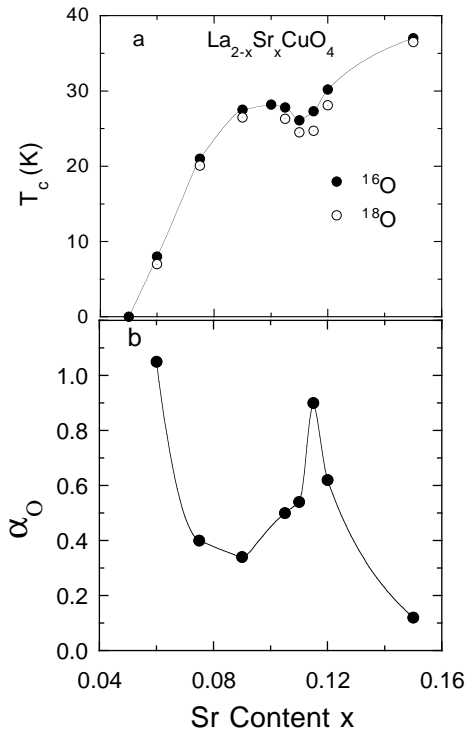}}
\caption[~]{Dependencies of $T_{c}$ and the oxygen-isotope exponent 
$\alpha_{O}$ on the Sr content $x$ for $^{16}$O 
and $^{18}$O samples in La$_{2-x}$Sr$_{x}$CuO$_{4}$. After 
\cite{ZhaoJPCM}.}
\label{IF1}
\end{figure}
Therefore, if the electron-phonon interaction is strong 
enough to form polarons and/or bipolarons, 
one will expect a substantial isotope effect on 
the effective mass of carriers. In this article, we will present 
various isotope effects in cuprates including the unconventional 
isotope effects on the supercarrier mass, on the charge-stripe 
formation temperature, on the pseudogap formation temperature, on 
the EPR linewidth, on the spin-glass freezing temperature, and on the 
antiferromagnetic ordering temperature. These unconventional isotope 
effects clearly demonstrate that phonons are relevant to the basic physics of 
cuprates and may be important for the occurrence of high-temperature 
superconductivity.

\section{Isotope effect on the superconducting transition temperature} 

Studies of isotope 
shifts of $T_{c}$ have been carried out in almost all 
known cuprates. A comprehensive review was given by Franck \cite{Franck94b}. 
The role of the anharmonicity of the apical oxygen on the isotope effect 
was discussed in detail by M\"uller \cite{Muller90}.
Most of the studies reported so far concern the oxygen-isotope shift 
(OIS) of $T_{c}$ by replacing $^{16}$O with 
$^{18}$O, partly because the experimental procedures are
simple and reliable. Now it is generally accepted that optimally-doped cuprates exhibit a 
small and positive oxygen-isotope exponent $\alpha_O$. It was also 
found that $\alpha_O$ for optimally-doped materials decreases with 
increasing $T_{c}$ \cite{SKBO}. The small OIS 
observed in the optimally doped cuprates suggest that phonons might 
not be important in bringing about high temperature superconductivity.
\begin{figure}[htb]
\input{epsf}
\epsfxsize7cm 
\centerline{\epsfbox{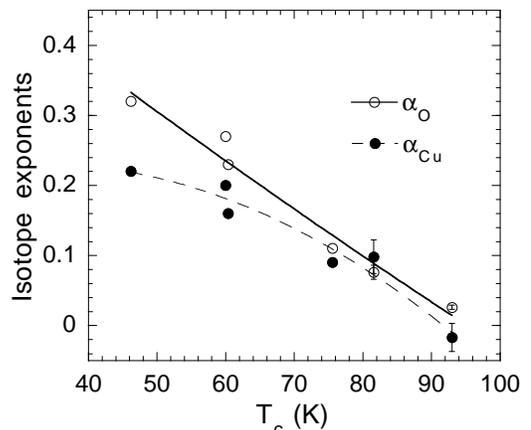}}
\caption[~]{The oxygen and copper 
isotope exponents as a function of $T_{c}$ for 
Y$_{1-x}$Pr$_{x}$Ba$_{2}$Cu$_{3}$O$_{7-y}$ and 
Y$_{1-x}$Pr$_{x}$Ba$_{2}$Cu$_{4}$O$_{8}$. The data were taken from 
\cite{ZhaoCu,Morris,Williams}.}
\label{IF2}
\end{figure}
However, the doping dependence of the OIS has been extensively studied 
in different cuprate systems  
\cite{Franck94b,Crawford90,Bornemann92,Franck93,Zech95,Zech96,ZhaoLSCO,ZhaoNature97,ZhaoJPCM}. For a particular family of doped cuprates the OIS increases with 
decreasing $T_c$, and can be even 
larger than the BCS value. In Fig.~\ref{IF1}a, we plot the doping 
dependence of $T_{c}$ for the $^{16}$O and $^{18}$O samples of the 
single-layer cuprate La$_{2-x}$Sr$_{x}$CuO$_{4}$. It is clear that 
the $T_{c}$'s of $^{18}$O samples are always lower than those of the $^{16}$O 
samples. The doping dependence of the isotope exponent $\alpha_O$ is 
shown in Fig.~\ref{IF1}b. The magnitude of $\alpha_O$ increases with a decrease 
of doping and becomes very large ($>$ 0.5) in the deeply underdoped 
regime. 
The large $\alpha_O$ value observed near $x$ = 0.125 might be related 
to the structural instability \cite{Pickett91}. The results suggest that the phonon 
modes related to the oxygen vibrations are strongly coupled to 
conduction electrons.
\begin{figure}[htb]
\ForceWidth{7cm}
\centerline{\BoxedEPSF{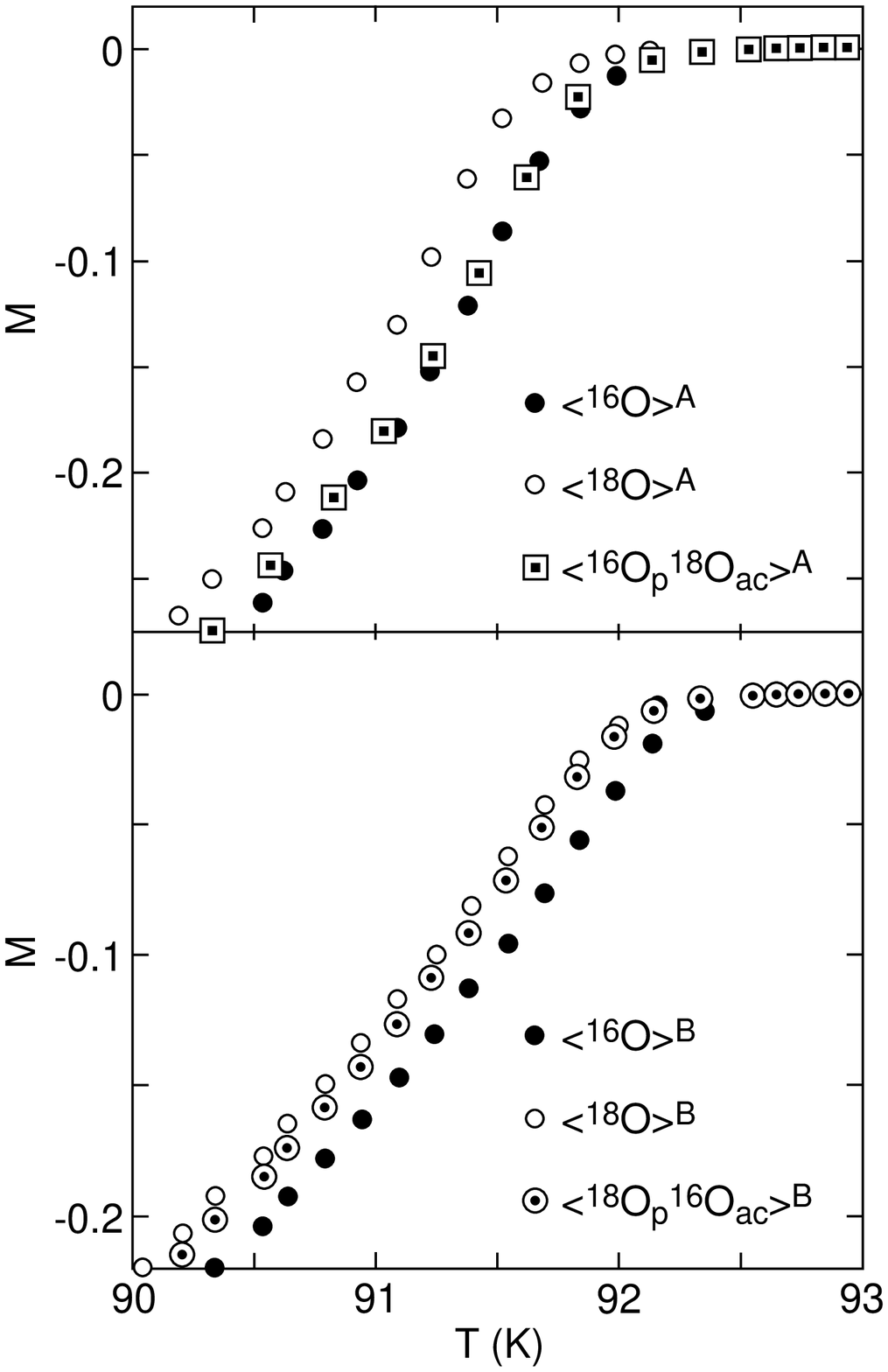}}
\caption[~]{The site-selective 
oxygen-isotope effect in YBa$_{2}$Cu$_{3}$O$_{6.96}$. Here 
$<$$^{18}$O$_{p}$$^{18}$O$_{ac}$$>$ means all the oxygen sites are 
exchanged by $^{18}$O, and $<$$^{18}$O$_{p}$$^{16}$O$_{ac}$$>$ means 
that only the in-plane oxygen sites are replaced by $^{18}$O. It is 
evident that the 
planar oxygen mainly 
($>80\%$) contributes to the total oxygen-isotope shift in the 
optimally doped cuprate. After 
\cite{ZechS}. }
\label{IFZ}
\end{figure}
In addition to the large oxygen-isotope shifts observed in underdoped 
compounds, there are also large 
copper isotope shifts observed in underdoped  
La$_{2-x}$Sr$_{x}$CuO$_{4}$ \cite{Franck93}, 
oxygen-depleted YBa$_{2}$Cu$_{3}$O$_{7-y}$ \cite{ZhaoCu}, Pr-substituted 
Y$_{1-x}$Pr$_{x}$Ba$_{2}$Cu$_{3}$O$_{7-y}$ \cite{Morris} and 
Y$_{1-x}$Pr$_{x}$Ba$_{2}$Cu$_{4}$O$_{8}$ \cite{Morris}, as well as in 
YBa$_{2}$Cu$_{4}$O$_{8}$ \cite{Williams}. In Fig.~\ref{IF2}, we plot the oxygen and copper 
isotope exponents as a function of $T_{c}$ in 
Y$_{1-x}$Pr$_{x}$Ba$_{2}$Cu$_{3}$O$_{7-y}$ and 
Y$_{1-x}$Pr$_{x}$Ba$_{2}$Cu$_{4}$O$_{8}$. As $T_{c}$ or doping 
decreases, both $\alpha_O$ and $\alpha_{Cu}$ increase monotonously. 
Interestingly, $\alpha_{Cu}$ is about 3/4 of $\alpha_O$ in the deeply 
underdoped region, while $\alpha_{Cu}$ is even larger than $\alpha_O$ 
near the optimal doping. This suggests that the Cu-dominated phonon 
modes (most of them are low-energy modes) are involved in the 
superconducting pairing. 
\begin{figure}[htb]
\ForceWidth{7cm}
\centerline{\BoxedEPSF{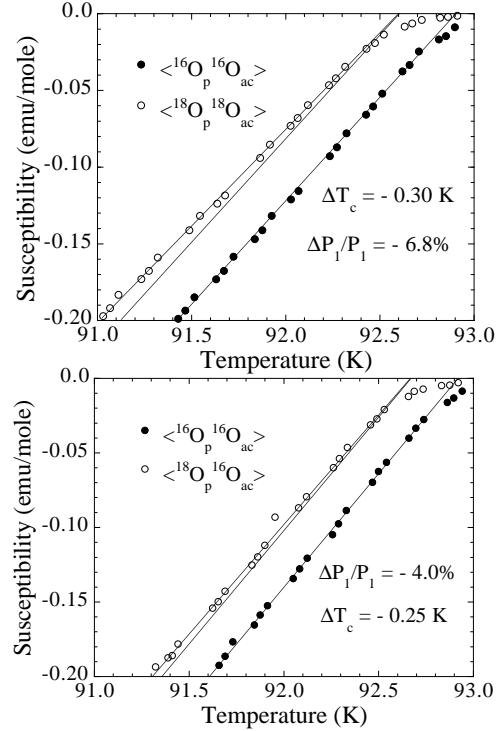}}
\caption[~]{The site-selective 
oxygen-isotope effect in an optimally-doped YBa$_{2}$Cu$_{3}$O$_{6.94}$.  After 
\cite{ZhaoS}. }
\label{IF3}
\end{figure}
The large copper-isotope shift also implies that the phonons in 
the CuO$_{2}$ planes are relevant to superconductivity. By analogy, 
one should also expect that the planar oxygen 
vibrations make more important contributions to the pairing than the 
apical and/or chain oxygen vibrations. We can distinguish the contributions of 
the different oxygen sites to the total OIS by site-selective 
oxygen-isotope experiments.  A partially site-selective oxygen-isotope 
exchange was attempted by Cardona {\em et al.} \cite{Cardona} and by 
Ham {\em et al.} \cite{Ham}.  The experimental difficulty in doing a 
complete site-selective oxygen-isotope exchange was overcome by Nickel 
{\em et al.} \cite{Nickel93}.  In their experiment the authors 
replaced the $^{18}$O by $^{16}$O in the chain and apical oxygen sites 
of a fully $^{18}$O exchanged YBa$_{2}$Cu$_{3}$O$_{7}$ sample, while 
keeping the $^{18}$O unexchanged in the plane sites.  The 
site-selectivity was confirmed by Raman spectroscopy \cite{Nickel93}.  
A small negative OIS [$\alpha_{O} = -0.010(4)$] associated with the 
planar oxygen was found.  In contrast, Zech {\em et al.} \cite{ZechS} 
showed a small positive OIS [$\alpha_{O} = 0.018(4)$] related to the 
planar oxygen ions.  The above discrepancy may be due to a broad 
superconducting transition in the samples of Nickel {\em et al.} 
\cite{Nickel93}, which makes it difficult to define $T_{c}$ reliably.

\begin{figure}[htb]
\ForceWidth{7cm}
\centerline{\BoxedEPSF{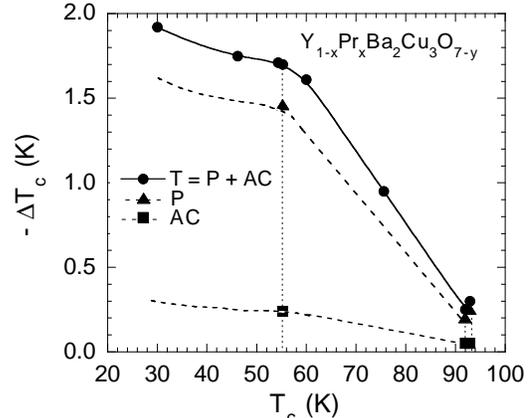}}
\caption[~]{The total (T) isotope shifts as a function of $T_{c}$ 
for Y$_{1-x}$Pr$_{x}$Ba$_{2}$Cu$_{3}$O$_{7-y}$ together with the 
isotope shifts from the planar (P) oxygen ions as well as from the apical and 
chain (AC) oxygen ions. The partial data of the total oxygen-isotope shifts are 
taken from Ref.~\cite{Franck94b} and have been corrected for 
100$\%$ oxygen-isotope exchange. The other data are from 
Refs.\cite{ZechS,ZhaoS}.}
\label{IFS}
\end{figure}

Although the OIS is small in the optimally-doped samples, the experiment 
done by Zech {\em et al.} \cite{ZechS} provided evidence 
that the planar oxygen ions mainly (80$\%$) contribute to 
the total OIS (see Fig.~\ref{IFZ}). Nevertheless it is not obvious that the same 
conclusion should apply to the underdoped samples where the total OIS is 
large. Zhao {\em et al.} \cite{ZhaoS} thus carried out the 
site-selective oxygen-isotope experiments in the underdoped and optimally 
doped samples of Y$_{1-x}$Pr$_{x}$Ba$_{2}$Cu$_{3}$O$_{7-y}$. 
~In Fig.~\ref{IF3}, ~~we 
\begin{figure}[htb]
\input{epsf}
\ForceWidth{7cm}
\centerline{\BoxedEPSF{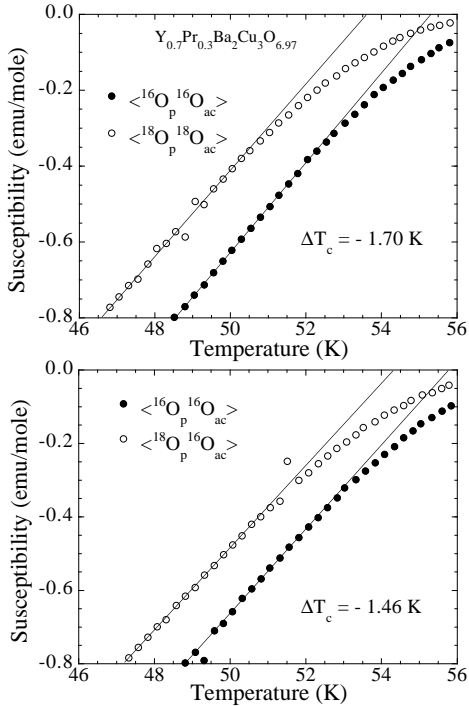}}
\caption[~]{The site-selective 
oxygen-isotope effect in 
an underdoped Y$_{0.7}$Pr$_{0.3}$Ba$_{2}$Cu$_{3}$O$_{6.97}$. The results show that the 
planar oxygen mainly 
($>80\%$) contributes to the total oxygen-isotope shift in this 
underdoped cuprate. After 
\cite{ZhaoS}.}
\label{IF4}
\end{figure}
\noindent
show the results of the site-selective oxygen-isotope effect 
for optimally doped YBa$_{2}$Cu$_{3}$O$_{6.94}$.  It is striking that 
the results shown in Fig.~\ref{IF3} are in excellent agreement with 
those in Fig~\ref{IFZ}.  This indicates a good reproducibility of 
these experiments.  The results for underdoped 
Y$_{0.7}$Pr$_{0.3}$Ba$_{2}$Cu$_{3}$O$_{6.97}$ are plotted in 
Fig.~\ref{IF4}.  Remarkably, the planar oxygen ions mainly ($>80\%$)
contribute to the total OIS in the underdoped
samples as well. In Fig.~\ref{IFS}, we show 
the total isotope shifts as a function of $T_{c}$ 
for Y$_{1-x}$Pr$_{x}$Ba$_{2}$Cu$_{3}$O$_{7-y}$ together with the 
isotope shifts from the planar oxygen ions as well as from the apical and 
chain oxygen ions.  From this figure, one can clearly see that the 
planar oxygen ions make a predominant contribution to the total OIS in all the doping 
levels.

\section{Negligible oxygen-isotope effect on the carrier density}

From the above results, an important question arises: Are the observed large isotope 
shifts in underdoped cuprates caused by a possible difference in 
the carrier densities of two isotope samples or by a strong 
electron-phonon coupling? It is very 
unlikely that the isotope effect is due to a difference in 
the carrier densities of two isotope samples. This is because 
the $T_{c}$'s of the $^{18}$O samples are always 
lower than the $^{16}$O samples by more than 1 K in the underdoped 
region, independent of whether $dT_{c}/dx$ is positive, negative, or 
zero (see Fig.~\ref{IF1}a).

\begin{figure}[htb]
\input{epsf}
\epsfxsize7cm 
\centerline{\epsfbox{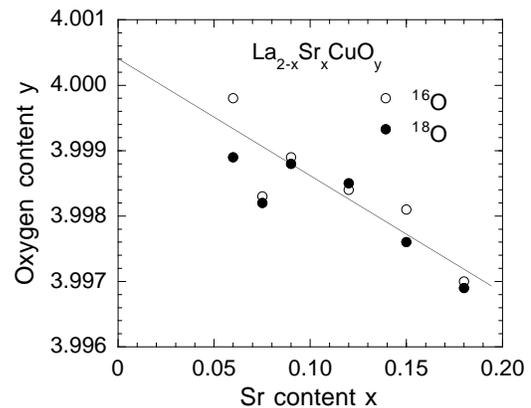}}
\caption[~]{The oxygen contents of the $^{16}$O and $^{18}$O samples 
of La$_{2-x}$Sr$_{x}$CuO$_{y}$ with different doping $x$ (unpublished 
data). The oxygen 
content was determined from a very precise volumetric analysis 
\cite{Conder}. The weight of each sample used for 
the analysis is about 500 mg. The oxygen content can be determined 
more precisely for a higher $x$ and a heavier sample. The oxygen 
contents of two isotope samples are the 
same within $\pm$0.0002 per Cu site.}
\label{IF5}
\end{figure}

There are three indirect experiments which have demonstrated that the 
difference in the hole densities of the $^{16}$O and $^{18}$O samples 
is smaller than 0.0002 per Cu site \cite{ZhaoLSCO,ZhaoNature97,ZhaoJPCM}. Now we 
are able to determine the 
oxygen content very accurately using a very precise volumetric
analysis \cite{Conder}.  Fig.~\ref{IF5} shows the oxygen contents 
of the $^{16}$O
\noindent
and $^{18}$O samples of La$_{2-x}$Sr$_{x}$CuO$_{4}$ with different doping 
$x$.  It is remarkable that the oxygen contents of two isotope samples 
are the same within $\pm$0.0002 per Cu site.  These experiments 
consistently show that the difference in the hole densities of the 
$^{16}$O and $^{18}$O samples is negligible, so that the observed 
large isotope effects are intrinsic and caused by a strong 
electron-phonon interaction.

\section{Large oxygen-isotope effect on the effective supercarrier 
mass}

Several groups \cite{Bornemann92,Franck93} noticed that there is an isotope effect on the 
diamagnetic Meissner signal in decoupled fine-grained samples. 
Zhao {\em et al.}\cite{ZhaoLSCO} carefully studied this effect and 
interpreted it as due to the isotope-mass dependence of the average 
supercarrier mass $m^{**}$ ($\equiv$ $\sqrt[3]{(m^{**}_{ab})^{2}m^{**}_{c}}$~).  
Since the magnetic 
penetration depth $\lambda(0)$ is proportional to 
$\sqrt{m^{**}/n_{s}}$, then
\begin{equation}\label{Ie3}
\Delta m^{**}/m^{**}= 2\Delta\lambda(0)/\lambda(0) + \Delta n_{s}/n_{s},
\end{equation}
where $\Delta$ means any small change of a quantity upon isotope 
substitution. Thus the isotope dependence of $m^{**}$ can 
be determined if one can independently measure the isotope dependence of 
$\lambda(0)$ and of $n_{s}$. As shown above, there is a negligible 
oxygen-isotope effect on the normal carrier density $n$, and $n_{s}$ 
= $n$ for clean superconductors, so one should expect no 
significant isotope effect on $n_{s}$. Indeed, there is a negligible 
oxygen-isotope effect on the supercarrier density $n_{s}$ in 
YBa$_{2}$Cu$_{3}$O$_{6.94}$ \cite{ZhaoYBCO}.

The isotope dependence of $\lambda(0)$ can be determined from that of 
the Meissner fraction $f(0)$ which, for decoupled and fine-grained 
samples, depends on the 
penetration depth $\lambda(0)$ and on the average grain radius $R$, as seen 
from the Shoenberg formula for spherical grains \cite{Shoenberg}:
\begin{equation}\label{Ie4}
f(T)= \frac{3}{2}[1- 3(\frac{\lambda(T)}{R})\coth (\frac{R}{\lambda(T)}) + 
3(\frac{\lambda(T)}{R})^{2}],
\end{equation}
where $\lambda(T)$ = 
$\sqrt[3]{[(\lambda_{ab}(T)]^{2}\lambda_{c}(T)}$ for layered compounds 
\cite{Kogan}. From equation \ref{Ie4}, it is obvious that a change in $\lambda(0)$ 
will lead to a change in $f(0)$, which implies that the isotope dependence of $\lambda(0)$ 
can be determined from the isotope dependence of $f(0)$.      

\begin{figure}[htb]
\input{epsf}
\epsfxsize7cm 
\centerline{\epsfbox{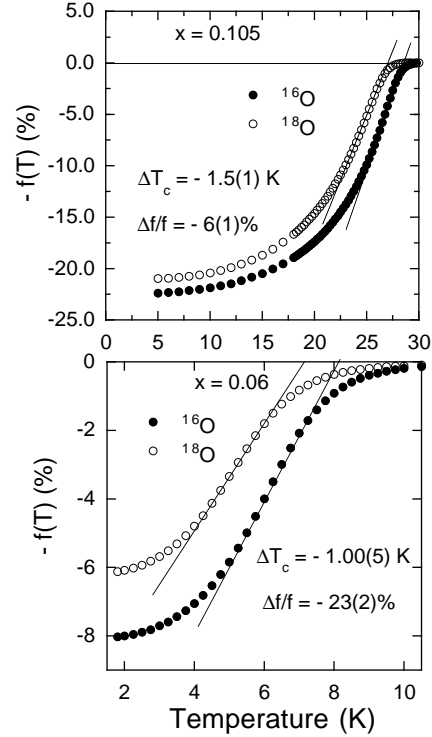}}
\caption[~]{The Meissner fractions for the $^{16}$O and $^{18}$O samples 
of La$_{2-x}$Sr$_{x}$CuO$_{4}$ with $x$ = 0.06 and 0.105. After 
\cite{ZhaoNature97,ZhaoJPCM}.}
\label{IF6}
\end{figure}

In Fig.~\ref{IF6}, we show the Meissner effects for the $^{16}$O and $^{18}$O samples 
of La$_{2-x}$Sr$_{x}$CuO$_{4}$ with $x$ = 0.06 and 0.105. The samples 
are loosely packed with rather small grain sizes ($R$ $\approx$ 
2-4 $\mu$m). One can clearly see that there are large oxygen isotope 
effects on both $T_{c}$ and the Meissner fraction. The most 
remarkable result is that the Meissner fraction of the $^{18}$O sample 
is lower than that for the $^{16}$O sample by about 23$\%$ in the case of 
$x$ = 0.06. The isotope effects are reversible upon the isotope 
back-exchange \cite{ZhaoNature97,ZhaoJPCM}, and reproducible in several 
sets of samples. For 
$x$ = 0.06, we evaluate $\Delta m^{**}/m^{**}$ = 24(2)$\%$. This 
indicates a large negative oxygen-isotope effect on the effective 
supercarrier mass in the deeply underdoped cuprates.

The above results were obtained from powder samples and 
correspondingly reflect the average properties of those highly anisotropic 
superconductors. To gain more insight, it is essential to determine the oxygen-isotope effect on the 
inplane effective supercarrier mass $m^{**}_{ab}$. Although we have 
tentatively extracted the isotope dependence of $m^{**}_{ab}$ from 
experiments on powder samples \cite{ZhaoNature97}, more reliable results should be obtained 
from experiments on single crystals.

Unfortunately, a complete oxygen-isotope exchange by gas diffusion is
impossible in single crystals with a large volume. In order to reach a complete 
oxygen-isotope exchange, microcrystals
with a volume of $V \approx 150 \times 150 \times 50 \ \mu$m$^{3}$
(mass $\approx 10 \ \mu$g) should be used. In this case, commercial SQUID 
magnetometers do not have enough sensitivity to measure the 
magnetization for such tiny crystals especially near $T_{c}$. 
Fortunately our highly sensitive torque magnetometer 
\cite{WilleminJAP98} is able to detect the small diamagnetic signal for the 
tiny crystals.

\begin{figure}[htb]
\input{epsf}
\epsfxsize7cm
\centerline{\epsfbox{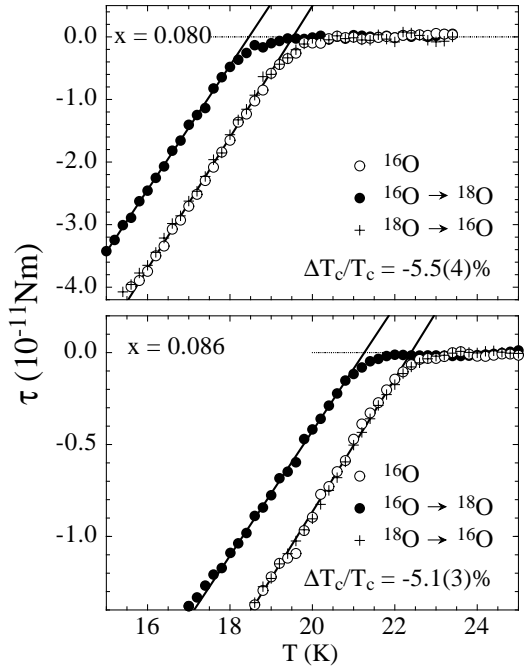}}
\caption[~]{The magnetic torque as a function of temperature for the $^{16}$O and 
$^{18}$O microcrystals of La$_{2-x}$Sr$_{x}$CuO$_{4}$ with $x$ = 
0.086 and 0.080.  The measurements were carried out in a 
magnetic field $B_{a} = 0.1$ T applied at an angle of  45$^{\circ}$ with 
respect to the c-axis. After 
\cite{HoferPRL}.}
\label{IF7}
\end{figure}

The superconducting transition was studied by cooling the sample in a magnetic
field $B_{a} = 0.1$ T applied at an angle of  45$^{\circ}$ with 
respect to the c-axis \cite{HoferPRL}. The torque signal
was continuously recorded upon cooling the sample at a
cooling rate of 0.01 K/s. In order to determine the
background signal of the cantilever, the measurement was repeated in
zero field and the data were subtracted from those of the 
field cooled measurement. The magnetic torque vs temperature 
data obtained for crystals with $x$ = 0.086 and 0.080 
are shown in Fig.~\ref{IF7}. It is clear that $T_{c}$ is lower for the $^{18}$O exchanged samples. 
Furthermore, the magnetic signals of the back-exchanged
samples (cross symbols) coincide with those of the
$^{16}$O annealed samples (open circles).
We define $T_{c}$ as the temperature where the linearly extrapolated
transition slope intersects the base line.
The relative changes
in $T_{c}$ are found to be $\Delta T_{c} / T_{c} = - 5.5(4)\%$ for 
$x = 0.080$ and  $-5.1(3)\%$ for $x$ = 0.086. 
The exponent $\alpha_{O}$ is found to be 0.47(2) for $x = 0.080$ and 0.40(2) for $x = 0.086$, which is in good 
agreement with the results obtained for powder samples with similar 
doping levels \cite{Crawford90,ZhaoJPCM}.

\begin{figure}[htb]
\input{epsf}
\epsfxsize7cm
\centerline{\epsfbox{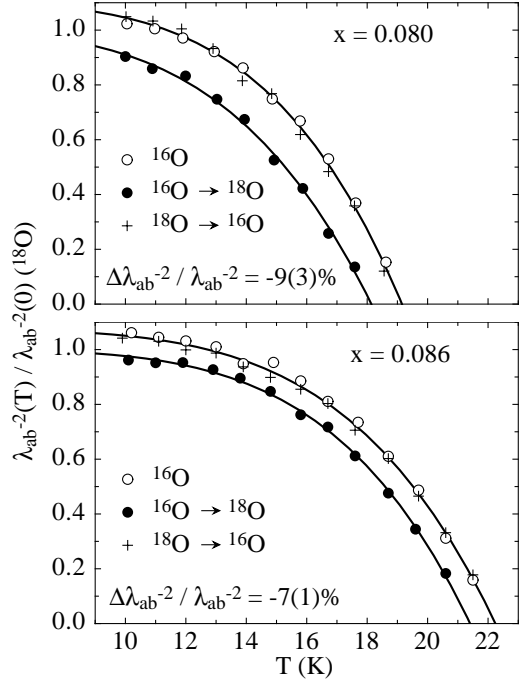}}
\caption[~]{The inplane penetration depth $\lambda_{ab}^{-2}(T)$ 
for the $^{16}$O and 
$^{18}$O microcrystals of La$_{2-x}$Sr$_{x}$CuO$_{4}$ with $x$ = 
0.086 and 0.080. The $\lambda_{ab}^{-2}(T)$ value was extracted from 
the field dependence of the torque.  After 
\cite{HoferPRL}.}
\label{IF8}
\end{figure}

More interestingly, the inplane penetration depth $\lambda_{ab}(T)$ 
can be extracted from field-dependent measurements \cite{HoferPRL}.
Fig.~\ref{IF8} displays
$\lambda_{ab}^{-2}(T)$ for the isotope-exchanged crystals 
with $x$ = 0.086 and 0.080. The temperature dependence is described by the power law
$\lambda_{ab}^{-2}(T) = \lambda_{ab}^{-2}(0)[1 - (T/T_{c})^{n}]$ with an
exponent $n \approx 5$. From Fig.~\ref{IF8} it is evident that both
$T_{c}$ and $\lambda_{ab}^{-2}(0)$ shift down upon replacing $^{16}$O by
$^{18}$O. The shifts are found to be
$\Delta \lambda_{ab}^{-2}(0) / \lambda_{ab}^{-2}(0) = -9(3)\%$ and
$-7(1)\%$ for $x = 0.080$ and
0.086, respectively. Using $\Delta n_{s}$ = 0, we find $\Delta 
m^{**}_{ab}/m^{**}_{ab}$ = 9(3)$\%$ for $x$ = 
0.080, and 7(1)$\%$ for $x$ = 0.086. 

Moreover, a substantial value of $\Delta m_{ab}^{**}/m_{ab}^{**}$ = 5-6$\%$ was 
also obtained for several optimally doped cuprates (e.g., YBa$_{2}$Cu$_{3}$O$_{6.94}$, 
La$_{1.85}$Sr$_{0.15}$CuO$_{4}$, and  
Bi$_{1.6}$Pb$_{0.4}$Sr$_{2}$Ca$_{2}$Cu$_{3}$O$_{10+y}$) from magnetization 
measurements \cite{Zhaounpub}. Very recent muon-spin rotation experiments on the 
oxygen-isotope exchanged YBa$_{2}$Cu$_{3}$O$_{6.96}$ have confirmed that $\Delta 
m_{ab}^{**}/m_{ab}^{**}$ $\simeq$ 5$\%$. Therefore, there is a 
substantial isotope effect on the in-plane supercarrier mass in 
optimally doped cuprates, but a very small effect on $T_{c}$. Such 
unusual isotope effects place strong constraints on the pairing 
mechanism of high-temperature superconductivity.

\section{Huge oxygen-isotope effect on the charge stripe formation 
temperature}

One of the most remarkable findings in the high-temperature copper 
oxide superconductors 
is the formation of alternating spin and charge stripes below a 
characteristic temperature \cite{Tranquada,Mook}. Various x-ray absorption 
spectroscopic measurements \cite{Bianconi92} suggest 
that the local structures in the alternating 
stripes are different, forming an incommensurate superlattice. 
Such a stripe phase is believed to be 
important to the understanding of the pairing mechanism of 
high-temperature superconductivity \cite{Emery}. However, 
the microscopic origin of 
the stripe phase is still highly debated. It could be caused by purely 
electronic interactions \cite{Emery} and/or by a strong electron-phonon 
interaction \cite{Holder}.

Although there is increasing experimental evidence for a strong 
electron-phonon interaction in the cuprate superconductors, it is not 
clear whether this interaction is important to the formation of the 
stripe phase. For the colossal magnetoresistive manganites, a strong 
electron-phonon interaction plays an essential role in the formation 
of the Jahn-Teller stripes (or charge ordering), as inferred 
from a 
very large oxygen-isotope shift of the charge-ordering temperature 
observed in 
both Nd$_{0.5}$Sr$_{0.5}$MnO$_{3}$ and La$_{0.5}$Ca$_{0.5}$MnO$_{3}$ 
systems \cite{ZhaoCO}.  Therefore it seems natural to seek for an 
isotope effect on the stripe formation temperature in cuprates.
\begin{figure}[htb]
\input{epsf}
\epsfxsize9cm
\centerline{\epsfbox{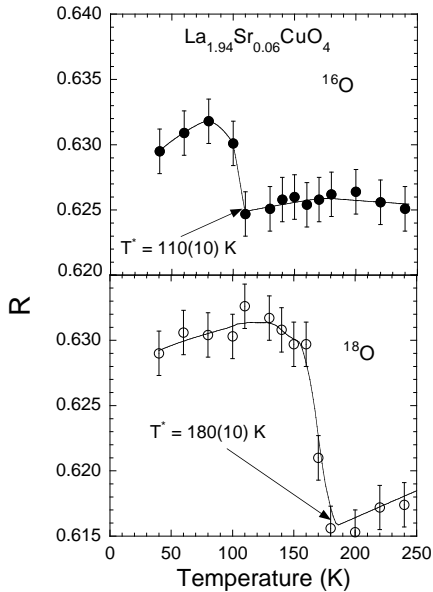}}
\caption[~]{The temperature dependence of the 
parameter $R$ for the oxygen-isotope exchanged 
La$_{1.94}$Sr$_{0.06}$CuO$_{4}$. See text for the definition of the 
parameter $R$. After 
\cite{Lanzara}.}
\label{IF9}
\end{figure}
X-ray absorption 
near edge spectroscopy (XANES) is a powerful technique to probe the 
local structure conformations for a system. From the XANES data, two 
characteristic peaks denoted by $A$ and $B$ in 
the XANES spectra  of the cuprates are identified, which 
characterize the local structures within and out of 
the CuO$_{2}$ planes.  A parameter $R$ is defined as 
$R = (\beta_{1}- \alpha_{1})/(\beta_{1}+ \alpha_{1})$, where 
$\beta_{1}$ and $\alpha_{1}$ are the intensities of peak $A$ and peak $B$ 
in the XANES spectra, respectively.  When charge-stripe ordering 
occurs, a change in the local 
structure takes place, leading to a sudden increase in $R$ below the 
charge-stripe formation temperature $T^{*}$. The identification of 
$T^{*}$ by XANES has been tested in a compound 
La$_{1.875}$Ba$_{0.125}$CuO$_{4}$ where $T^{*}$ was determined by 
other techniques \cite{Lanzara}.  
In Fig.~\ref{IF9}, we show the temperature dependence of the 
parameter $R$ for the oxygen isotope exchanged 
La$_{1.94}$Sr$_{0.06}$CuO$_{4}$.  From the 
figure, one can clearly see that, upon replacing 
$^{16}$O with $^{18}$O,  the 
charge-stripe formation temperature $T^{*}$ in this cuprate 
increases from about 110 K to 180 K. Such a large negative isotope 
effect on charge ordering can be explained in terms of 
electron-phonon coupling beyond the Migdal approximation \cite{Millis1}.

\section{Oxygen-isotope effects on the pseudogap formation temperature}
\begin{figure}[htb]
\input{epsf}
\epsfxsize 7cm
\centerline{\epsfbox{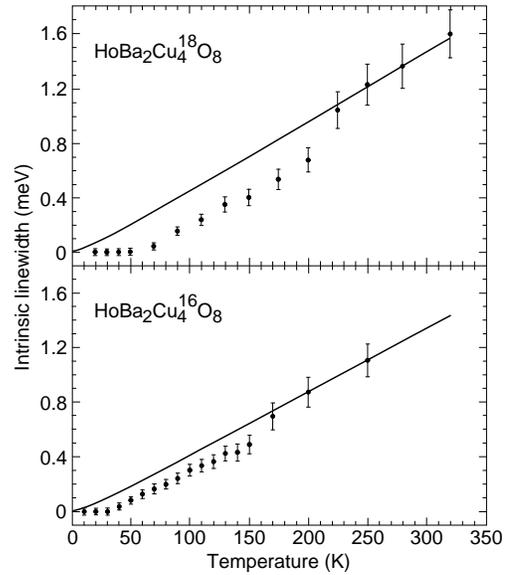}}
\caption[~]{Temperature dependence of the intrinsic linewidth $\Gamma 
(T)$ of the $\Gamma_{3}^{(1)}$ $\rightarrow$ $\Gamma_{4}^{(1)}$ transition 
of Ho$^{3+}$ for the $^{16}$O and $^{18}$O samples of 
HoBa$_{2}$Cu$_{4}$O$_{8}$. After Ref.~\cite{Furer}.}
\label{neutron}
\end{figure}
Recently, a neutron 
spectroscopic investigation of the isotope effect on the relaxation 
rate of crystal field excitations of Ho$^{3+}$ in 
HoBa$_{2}$Cu$_{4}$O$_{8}$ has 
been carried out \cite{Furer}. In Fig.~\ref{neutron}, we plot the 
temperature dependence of the intrinsic linewidth $\Gamma (T)$ (corresponding 
to the $\Gamma_{3}^{(1)}$ $\rightarrow$ $\Gamma_{4}^{(1)}$ transition 
of Ho$^{3+}$) for the $^{16}$O and $^{18}$O samples. In the high temperature 
range, the linewidth appears to exhibit a linear temperature dependence. Cooling 
down to a characterisitc 
temperature $T^{p}$, the linewidth suddenly gets narrower. Of 
particular interest is the characterisitc 
temperature $T^{p}$ which strongly depends on the oxygen-isotope mass. Upon  
replacing $^{16}$O by $^{18}$O, $T^{p}$ changes from 170 K to 220 K, 
i.e., there is a large oxygen-isotope 
effect on $T^{p}$ ($\approx$ 50 K). Similarly, a large copper-isotope 
effect on $T^{p}$ has been found in the same compound \cite{Furer1}. The huge 
oxygen-isotope effects
on $T^{*}$ \cite{Lanzara} and on $T^{p}$ \cite{Furer} indicate that  
a strong electron-phonon interaction plays an essential role in the 
charge dynamics and superconductivity \cite{Holder}. Although the 
authors of Ref.~\cite{Furer} attributed the characterisitc 
temperature $T^{p}$ to a pseudogap formation temperature, one cannot 
rule out the possibility that $T^{p}$ may be related to a dynamical 
charge ordering. We would like to mention that a small positive 
oxygen-isotope effect on the spin pseudogap has been observed in 
YBa$_{2}$Cu$_{4}$O$_{8}$ \cite{Raffa}.

\section{Oxygen-isotope effect on the EPR linewidth}
\begin{figure}[htb]
\input{epsf}
\epsfxsize 7cm
\centerline{\epsfbox{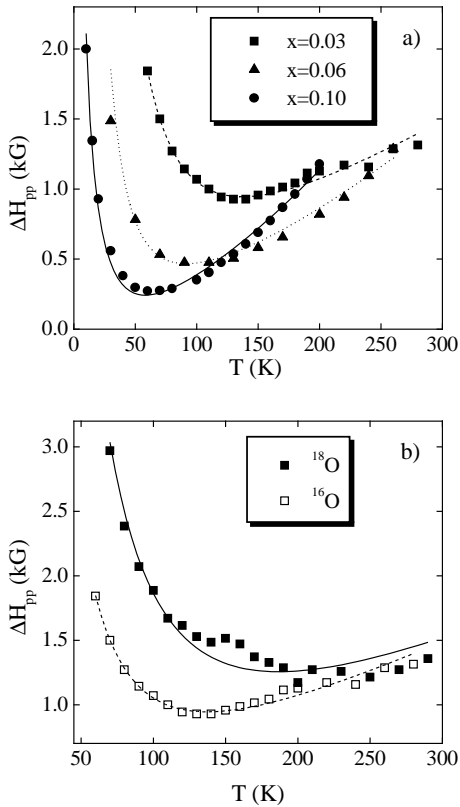}}
\caption[~]{(a) Temperature dependence of the  EPR linewidth $\Delta H_{pp}$ 
for $^{16}$O isotope samples of La$_{2-x}$Sr$_{x}$Cu$_{0.98}$Mn$_{0.02}$O$_{4}$
with $x=0.03$,  $0.06,$ and $0.10$; (b) The oxygen-isotope effect on 
the EPR linewidth $\Delta H_{pp}$ in 
La$_{1.97}$Sr$_{0.03}$Cu$_{0.98}$Mn$_{0.02}$O$_{4}$. After 
Ref.~\cite{Muller2000,Shengelaya}.}
\label{EPR}
\end{figure} 
The Electron Paramagnetic  Resonance (EPR) is a powerful technique to 
probe and understand various phenomena in solid state physics.  
Unfortunately, the intrinsic Cu$^{2+}$ signals in the cuprate superconductors 
have not been observed except for those of chain fragments in 
YBa$_{2}$Cu$_{3}$O$_{6+\delta}$ (0.7 $<$ $\delta$ $<$ 0.9) \cite{Sichelschmitt}, 
and three-spin polarons \cite{Kochelaev1}. The EPR silence has been 
attributed by many researchers to the very fast relaxation time of 
the Cu$^{2+}$ spins. In order to test this possibility, one needs to 
dope the material with a paramagnetic S-state 
ion (e.g.,  Mn$^{2+}$). Its relaxation occurs via the fast relaxing 
carriers. This is called bottleneck effect, which has indeed been observed in Mn$^{2+}$ doped La$_{2-x}$Sr$_{x}$CuO$_{4}$ \cite{Kochelaev2}.

Here we focus on the oxygen-isotope effect on the EPR linewidth in the 
Mn$^{2+}$ doped 
La$_{2-x}$Sr$_{x}$CuO$_{4}$ \cite{Muller2000,Shengelaya}. In Fig.~\ref{EPR}, we 
show the temperature 
dependence of the EPR linewidth $\Delta H_{pp}$ 
for $^{16}$O isotope samples of La$_{2-x}$Sr$_{x}$Cu$_{0.98}$Mn$_{0.02}$O$_{4}$
with $x=0.03$, $0.06,$ and $0.10$ (Fig.~\ref{EPR}a), as well as the 
isotope effect on 
the EPR linewidth $\Delta H_{pp}$ in 
La$_{1.97}$Sr$_{0.03}$Cu$_{0.98}$Mn$_{0.02}$O$_{4}$ (Fig.~\ref{EPR}b). It is remarkable 
that the EPR linewidth at 100 K for the $^{16}$O sample is about half the 
one for  
the $^{18}$O sample. Such a large oxygen-isotope effect on the 
linewidth indicates that the 
dynamics of the oxygen atoms play a key role in the Cu$^{2+}$ 
relaxation. Interestingly, the temperature dependencies of the 
linewidth for both isotope samples are in good agreement with 
theoretical calulations (see solid and dashed lines) \cite{Muller2000,Shengelaya}.

\section{Large oxygen-isotope effect on spin-glass freezing temperature}
\begin{figure}[htb]
\input{epsf}
\epsfxsize7cm
\centerline{\epsfbox{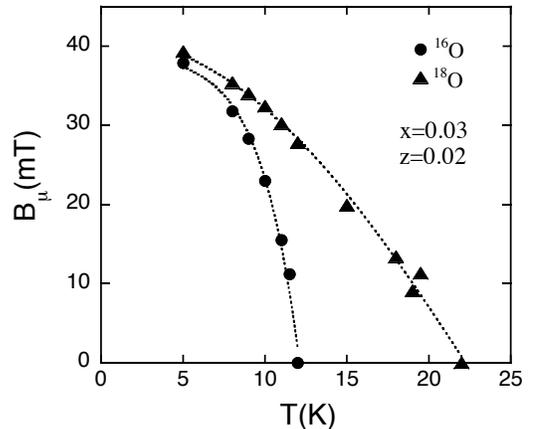}}
\caption[~]{Temperature dependence of the internal 
magnetic fields {\em B$_{\mu}$} (probed by muon-spin-rotation) for 
the $^{16}$O and $^{18}$O samples 
of La$_{1.97}$Sr$_{0.03}$Cu$_{0.98}$Mn$_{0.02}$O$_{4}$. After 
\cite{SG}.}
\label{IF10}
\end{figure}
The parent compounds of the cuprate
superconductors exhibit long-range 3D antiferromagnetic (AF) order,
which is rapidly destroyed as holes are doped into the  CuO$_{2}$ planes.
A short range ordered AF state exists at doping of 0.02$<$$x$$<$0.06 
in La$_{2-x}$Sr$_{x}$CuO$_{4}$. Early $\mu$SR
and neutron scattering experiments found that this magnetic state
resembles a spin glass \cite{mSR}. More detailed studies using
$^{139}$La nuclear quadrupole resonance \cite{NQR} showed that the 
magnetic state
in this doping regime is not a conventional spin
glass, but a cluster spin glass (CSG). The understanding of how 
the short-range ordered 
AF state affects superconductivity and how it is influenced by lattice 
vibrations will help to clarify the pairing mechanism of high-T$_{c}$
superconductivity.

It is well known that conventional theories of magnetism neglect atomic vibrations;
the atoms are mostly considered as being infinitely heavy in
theoretical descriptions of magnetic phenomena, which rules out the 
existence of an isotope effect on magnetism. However, there is an 
exceptional example. In the ferromagnetic manganites, a giant oxygen-isotope 
effect on the Curie temperature has been observed \cite{ZhaoNature96}.  
The question is whether such an isotope effect also exists in 
cuprates.

In Fig.~\ref{IF10}, we show the temperature dependence of the internal 
magnetic fields {\em B$_{\mu}$} 
(probed by muon-spin rotation) for 
the $^{16}$O and $^{18}$O samples 
of La$_{1.97}$Sr$_{0.03}$Cu$_{0.98}$Mn$_{0.02}$O$_{4}$. It is 
remarkable that the spin-glass freezing temperature $T_{g}$ almost 
doubles upon replacing $^{16}$O by $^{18}$O. Such a huge isotope 
effect on the spin-glass freezing temperature suggests that spin 
dynamics in cuprates is ultimately correlated with lattice vibrations.

\section{Oxygen-isotope effect on the AF ordering temperature}

The antiferromagnetic order observed in the parent insulating 
compounds like La$_{2}$CuO$_{4}$ signals a strong electron-electron 
Coulomb correlation.  On the other hand, the large isotope effects found 
in the underdoped 
cuprate superconductors indicate a strong electron-phonon 
interaction. Now a question 
arises: can the strong electron-phonon interaction modify the 
antiferromagnetic exchange energy and thus the AF ordering 
temperature in the parent insulating compounds? Studies of the 
isotope effect on the AF ordering temperature could clarify 
this issue \cite{ZhaoAF,HolderAF}.

\begin{figure}[htb]
\input{epsf}
\epsfxsize7cm
\centerline{\epsfbox{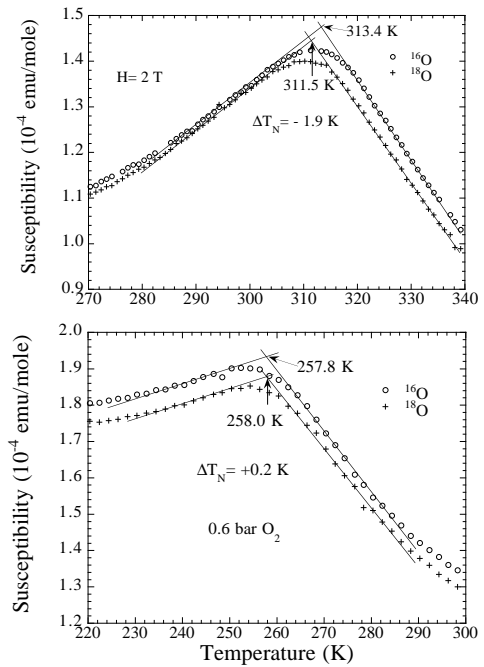}}
\caption[~]{The temperature dependence of the 
susceptibility for the $^{16}$O and $^{18}$O samples of undoped 
La$_{2}$CuO$_{4}$ (upper panel), and of the oxygen-doped 
La$_{2}$CuO$_{4+y}$ (lower panel).  After 
\cite{ZhaoAF}.}
\label{IF11}
\end{figure}

Fig.~\ref{IF11} shows the temperature dependence of the 
susceptibility for the $^{16}$O and $^{18}$O samples of undoped 
La$_{2}$CuO$_{4}$ (upper panel), and of oxygen doped 
La$_{2}$CuO$_{4+y}$ (lower panel). 
One can see that the AF ordering temperature $T_{N}$ for the $^{18}$O 
sample is lower than the $^{16}$O sample by about 1.9 K in the case 
of the undoped samples. For the oxygen-doped samples, there is a 
negligible isotope effect.

It is known that the antiferromagnetic properties of 
La$_{2}$CuO$_{4+y}$ can be well understood within 
mean-field theory which leads to a $T_{N}$ formula 
\cite{Thio}:
\begin{equation}\label{Ie5}
k_{B}T_{N} = J' [\xi(T_{N})/a]^{2},
\end{equation}
where $J'$ is the interlayer coupling energy, $\xi (T_{N})$ is the 
in-plane AF correlation length at $T_{N}$ with 
$\xi (T_{N}) \propto \exp (J/T_{N})$ for $y$ = 0 ($J$ is the in-plane 
exchange energy). When $T_{N}$ is 
reduced to about 250 K by oxygen doping, a mesoscopic phase 
separation has taken place so that $\xi (T_{N})$ = $L$ (Ref.~\cite{Cho}), where $L$ is 
the size of the antiferromagnetically correlated clusters, and depends 
only on the extra oxygen content $y$. In this case, we have $T_{N} = 
J'(L/a)^{2}$. Since $L$ is independent of the isotope mass, a negligible 
isotope shift of $T_{N}$ in the oxygen-doped La$_{2}$CuO$_{4+y}$ suggests that 
$J'$ is independent of the isotope mass. Then we easily find for 
undoped compounds
\begin{equation}\label{Ie6}                                       
\Delta T_{N}/T_{N} = (\Delta J/J)\frac{B}{1+B},
\end{equation}
where $B = 2J/T_{N} \simeq$ 10. From the measured isotope 
shift of $T_{N}$ for the undoped samples, we obtain $\Delta J/J$ $\simeq -0.6\%$.

Theoretically, Kugel and Khomski \cite{Kugel} considered the 
Jahn-Teller effect in 
a single-band Hubard model.  They showed 
that, for $U$ $>$$>$ $E_{p}$, $\hbar\omega$,
\begin{equation}\label{Ie7}
J = \frac{2t^{2}}{U}(1 + 
\frac{2E_{p}\hbar\omega}{U^{2}}).
\end{equation}
On the other hand, when $U$ $<$$<$ $\hbar\omega$,
\begin{equation}\label{Ie8}
J = \frac{2t^{2}\exp(-2E_{p}/\hbar\omega)}{U-2E_{p}}.
\end{equation}
~\\
Here $U$ is the onsite Coulomb repulsion, $t$ is the bare hopping 
integral, $E_{p}$ is the Jahn-Teller 
stabilization energy, and $\omega$ is the vibration frequency of the 
Jahn-Teller mode. Assuming $t$ = 0.5 eV, $J$ = 0.13 eV, we obtain 
$U$ = 3.8 eV. With $\hbar\omega$ = 0.1 eV, 
$E_{p}$ = 1.2 eV \cite{Kamimura}, we
obtain $\Delta J/J$ $\simeq -0.1\%$ from Eq.~\ref{Ie7}, which is about 
a factor of 6 smaller than the measured value. The discrepancy might 
be due to the fact that the single-band Hubard model is 
oversimplified for the parent cuprates.

\section{Concluding remarks}

In summary, the unconventional isotope effects observed in cuprates 
clearly demonstrate that the electron-phonon interaction plays an 
important role in the physics of cuprates. Our results also show that 
the phonon modes 
related to both oxygen and copper 
vibrations are important to the pairing.

Interestingly, the concept of enhancing the 
electron-phonon coupling  was the original motivation for the 
high-$T_c$ discovery.
Bednorz and M\"uller \cite{KAM86} argued that there should be 
a strong electron-phonon interaction in perovskites 
with strong Jahn-Teller centers. Indeed, the stretching 
vibration mode, which is related to the $Q_{2}$-type Jahn-Teller 
distortion, has proved to couple strongly to the doped 
holes \cite{Pint,McQueeney,Petrov}. This high-energy $Q_{2}$-type mode 
may couple to the 
low-energy tilting mode ($Q_{4}$/$Q_{5}$-like Jahn-Teller mode) if the 
Cu-O-Cu bonding angle is less than 180$^{\circ}$. Thus 
both stretching and tilting modes are
important to high-$T_{c}$ 
superconductivity.
~\\
~\\
~\\
{\bf Acknowlegement}: The authors would like to thank 
K. A. M\"uller, A. Shengelaya, J. Hofer, A. Bianconi and D. E. Morris for their 
collaborations and discussions. The work was supported by the Swiss 
National Science Foundation.

\end{document}